\documentclass[journal,comsoc]{IEEEtran}

\ifCLASSINFOpdf

\else

\fi

\usepackage{mathrsfs}
\usepackage{amsmath}
\usepackage{amsfonts}
\usepackage{amsthm}
\usepackage{amssymb}
\usepackage{graphicx}
\usepackage{subfigure}
\usepackage{indentfirst}
\usepackage{array}
\usepackage{epstopdf}
\usepackage{cite}
\usepackage{enumerate}
\usepackage{bm}
\usepackage{dsfont}
\usepackage[linesnumbered,ruled,vlined]{algorithm2e}
\usepackage{multirow}
\usepackage{verbatim}
\usepackage{stfloats}
\usepackage{color}

\SetKwComment{Comment}{//}{}
\SetKwBlock{Begin}{Function}{end}

\begin{document}
	
	\title{Generalized Polarization Transform: A Novel Coded Transmission Paradigm}
	
	\author{
		Bolin~Wu,
        Jincheng~Dai,
		Kai~Niu,
		Zhongwei~Si,
		Ping~Zhang,
		Sen~Wang,
		Yifei~Yuan,~and
		Chih-Lin I
		
        \thanks{\emph{Corresponding author: Jincheng Dai.}}
		\thanks{J. Dai, K. Niu, and Z. Si are with the Key Laboratory of Universal Wireless Communications, Ministry of Education, Beijing University of Posts and Telecommunications, Beijing 100876, China (e-mail: daijincheng@bupt.edu.cn; niukai@bupt.edu.cn).}
		\thanks{P. Zhang is with the State Key Laboratory of Networking and Switching Technology, Beijing University of Posts and Telecommunications, Beijing 100876, China.}
		\thanks{S. Wang, Y. Yuan, and C. I are with the China Mobile Research Institute, Beijing, China.}
		
		\vspace{-1em}
	}
	
	\maketitle	
	
	\begin{abstract}
		For the upcoming 6G wireless networks, a new wave of applications and services will demand ultra-high data rates and reliability. To this end, future wireless systems are expected to pave the way for entirely new fundamental air interface technologies to attain a breakthrough in spectrum efficiency (SE). This article discusses a new paradigm, named generalized polarization transform (GPT), to achieve an integrated design of coding, modulation, multi-antenna, multiple access, etc., in a real sense. The GPT enabled air interface develops far-reaching insights that the joint optimization of critical air interface ingredients can achieve remarkable gains on SE compared with the state-of-the-art module-stacking design.
	\end{abstract}
	
	\IEEEpeerreviewmaketitle
	
	\section{Introduction}\label{section_introduction}
	
	With the proliferation of various burgeoning applications, e.g., extended reality (XR) services, real-time interactive services, tele-medicine services, etc., the rising technical requirements will make the fifth-generation (5G) wireless systems struggling in future 2030+ \cite{Oulu_6G,Zhang_6G}. Therefore, a newer and upgraded generation of wireless communication systems equipped with spectrum-efficient transmission techniques is highly anticipated. Extremely high data rates and reliability call for a rethinking of current air interface techniques in favor of the growing demand for spectrum efficiency (SE). Though some existing technologies, such as massive multiple-input multiple-output (MIMO), can achieve decent SE, the gap to the performance limit of point-to-point link communications can still be narrowed. For future wireless systems, say beyond-5G (B5G), to design and implement air interface modules from a new perspective for seeking a breakthrough in SE should be seriously considered. Instead of conservative separate modular optimization, integrating different modules, e.g., channel coding, modulation, multi-antenna, etc., in a collaborative way will be a promising coded transmission paradigm on which we focus in this article.
	
	The demand for SE enhancement poses two main challenges in air interface technique optimization. On the one hand, almost all current air interface modules are designed based on the separation principle \cite{Shannon}, and each module itself has already been highly optimized to approach its theoretical limits. However, the optimality of the separation principle holds only under unlimited delay and complexity, whereas these idealistic assumptions cannot be met in practice. Hence, the state-of-the-art modular design may obstruct further optimization. On the other hand, though many efforts have been devoted to the joint design of multiple modules, the lack of a mighty guiding theory hinders the optimization of air interface techniques supporting future wireless systems. As Bertalanffy emphasized in his general system theory\cite{Ber_system}, ``the whole is greater than the sum of its parts''. A drawback of popular joint design currently lies in that it is only constrained to the transmitter or receiver end rather than transceiver coordination. Hence, a guiding theory that runs through multiple modules is needed to exploit potential performance improvement. Notably, in 1982, Ungerboeck proposed the trellis coded modulation (TCM) scheme \cite{TCM} that can achieve error performance improvement without sacrificing data rates or requiring more bandwidths by increasing free Euclidean distance. This pioneering work is the first successful attempt of integrated design merging modulation and coding. In this article, guided by the polarization principle described below, we consider a broader integrated design of air interface modules from a new perspective.
	
	Recently, polar codes have emerged as a breakthrough in coding theory \cite{Arikan}. This new code family actually benefits from an elegant effect named channel polarization. In the polarization procedure, the same raw channels diverge into two kinds of synthesized channels, whose reliability shows slight difference, after a single step of channel transform. We classify this transform as \emph{coding polarization (CP)}. By recursively applying CP transform over those resulting channels, the reliability of synthesized channels gradually shows the striking difference: the good ones get better, and the bad ones get worse. One can naturally transmit information bits over those reliable channels while assigning frozen bits to noisy ones. Actually, the theoretical rationale of CP stems from the chain rule of mutual information \cite{KaiNiu}. However, this chain rule is not constrained to the binary coding field only. In other words, the channel polarization phenomenon can be found actually universal in many other signal processing scenarios \cite{Seidl,Dai_mimo,Dai_noma}, e.g., constellation modulation, MIMO, etc. In these cases, multiple data streams are coupled at the receiver and successive cancellation (SC) detection is implemented. With the decoupling procedure, the reliability of different data streams gradually differentiates, just similar to CP. We call this effect as \emph{signal polarization (SP)}. Although there have been many advanced polarization-based physical layer solutions, most existing works only focus on some specific technique optimization. How to efficiently apply polarization transform, across multiple air interface modules, as a powerful tool to realize an integrated design remains an open research issue.
	
	In this article, we first summarize challenges in the optimization of coded systems. Then we propose the generalized polarization transform (GPT) as a joint description of CP and SP. Specifically, GPT is a fusion design of communication systems that consider both transmitter and receiver, with the polarization principle as the top-level instruction. For those non-binary modules, SP comes into play, and the original data is split into streams with varying reliability. On this basis, CP acts on the resulting data streams to manipulate binary digits. In a nutshell, the modules under the GPT framework are precisely matched and tightly coupled, which all follow the polarization backbone, with the ultimate goal of maximizing the polarization effect.
	
	\section{Challenges in Air Interface Optimization}\label{section_challenges}
	
	Compared to 5G, future wireless networks, say 6G, are spurred to achieve superior performance and have more stringent requirements \cite{Oulu_6G,Zhang_6G}. For example, the peak data rate should reach 100 Gbps to 1 Tbps, while for 5G it is 20 Gbps, and the SE should increase by 3--5 times. Besides, the transmission reliability should be at least six ``9'' to ensure critical applications. Confronted with these challenges, new air interface and transmission technologies appear essential, including new waveforms, channel coding, multi-antenna, etc., and a proper combination of these diverse techniques.
	
	Recall that one of the fundamental results in information theory is Shannon's separation theorem \cite{Shannon}, which describes the optimality of transmitting a source signal over a noisy channel by independently optimizing source coding and channel coding modules. This modular separation has had tremendous impacts from both practical and conceptual standpoints since the binary digits provide a standard type interface. In this way, the design of channel coding doesn't need to know the details of other signal processing modules and vice versa. However, despite its huge impact, the optimality of this modular design principle holds only under unlimited blocklength and complexity; and even under these assumptions, it breaks down in multi-point communications, e.g., multiuser scenarios, or for non-ergodic channel distributions.
	
	Regardless of its suboptimality, this ``divide-and-conquer'' approach is popular in the current air interface module design. It is not only due to the convenience of modularity but the lack of a forceful theory to derive the collaborative design with reasonable complexity. In current systems, different channel codes generally have their specialties, such as LDPC codes for data transmission \cite{Richardson_LDPC} and polar codes for control information transmission \cite{Bioglio_polar} in 5G. Also, quadrature amplitude modulation (QAM) with Gray labeling is invariably designated as the constellation modulation scheme. The MIMO/multiuser optimization has focused mainly on the beamforming/codebook design under the capacity maximization criterion, not considering the channel coding behavior under finite blocklengths. Under this separate design paradigm, every module has been highly optimized to approach its theoretical performance limit. Nonetheless, the system performance under a simple combination of these modules is still far from the Shannon limit, just like an awkward situation where ``$1 + 1 < 2$''. Consequently, the joint optimization of air interface modules under a finite blocklength becomes essential for further improving SE and transmission reliability.
	
	Joint processing is not a new subject. Since the advent of Turbo codes, the Turbo paradigm has been rapidly promoted to many other signal processing scenarios, for example, iterative multiuser detection \cite{Wang_iterative}. Turbo-based approaches exploit the extrinsic information given by one constituent as the \emph{a priori} information to assist others, formulating an iterative structure between signal processing and coding modules. In this way, the Turbo principle has been proven capable of approaching the maximum a posteriori (MAP) detection and decoding. However, there are limitations. For example, most Turbo-based approaches only operate at the receiving end, neglecting a collaborative design at the transmitting end. That results in a clear gap to the Shannon limit and means a potential performance improvement room. Furthermore, Turbo-based approaches are always iterative that causes additional delay and complexity. Consequently, new methods are required to attain a real sense of integrated design of coded transmission modules to enhance the performance in wireless systems.
	
	
	\section{The GPT Principle}\label{section_GPT}
	
	As the most sparkling breakthrough in coding theory over the past two decades, the discovery of polar codes re-energized the coding community and provided them a radically different way of approaching the channel capacity \cite{Arikan}. Despite being a relatively immature coding scheme, polar codes have been proven a fierce competitor to Turbo and LDPC codes. Hence, polar codes have already been selected as the coding scheme for control channels in the 5G standard.

\begin{figure}[htbp]
		\centering{\includegraphics[scale=0.47]{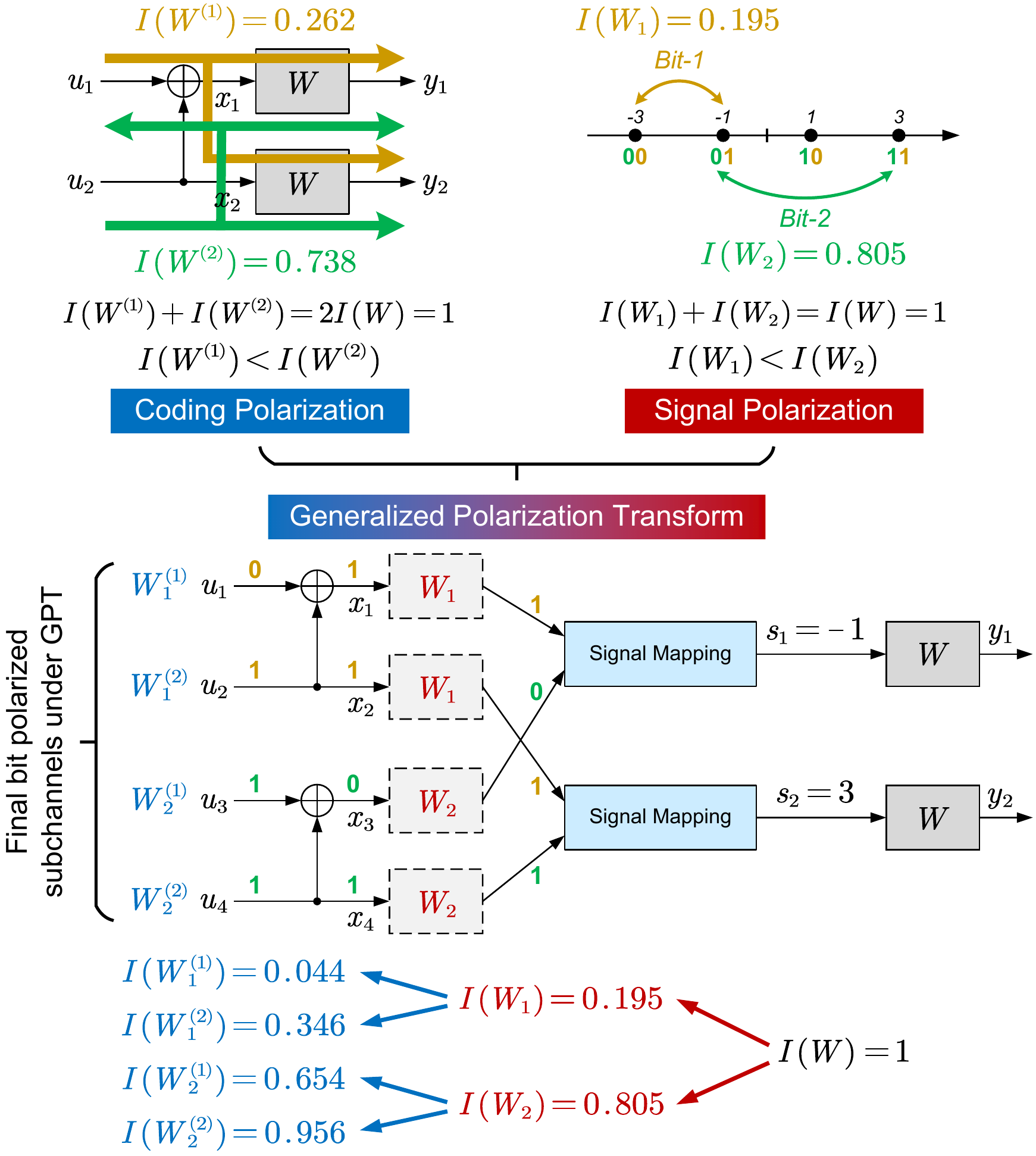}}
		\caption{A toy example of coding polarization, signal polarization, and generalized polarization transform.}
		\label{fig_1}
	\end{figure}

	The fundamental principle in polar codes is channel polarization transform, which stems from the chain rule of mutual information \cite{KaiNiu}. Actually, as long as multiple data streams are coupled, the successive detection/decoding acting as the channel splitting operation will bring reliability differentiation among data streams. Each term in the chain rule of mutual information corresponds to a split subchannel. In this way, the original channel can be decomposed as a set of subchannels with no capacity loss. Fig.~\ref{fig_1} provides a toy example to illustrate CP, SP, and GPT, respectively, to help the readers catch on to the process. In the figure, $I( \cdot )$ denotes the capacity of a symmetric channel, which measures the reliability. Obviously, the more reliable channels have greater capacities. Then, we describe each concept in the following subsections.
	
	\subsection{Coding Polarization}
	
	A single step of channel polarization transform is shown in Fig.~\ref{fig_1}, through which two raw channels $W$ are transformed into two synthesized channels $W^{\left(1\right)}$ and $W^{\left(2\right)}$, and their capacities are different while their sum preserves. This polarization transform in the binary field is summarized as CP. Polar codes are theoretically capacity-achieving with infinite blocklengths, but they cannot fully polarize the underlying channels in the finite blocklength regime. In practice, after channel combining and splitting, the raw channels are transformed into good and bad subchannels with clearly different reliabilities rather than completely noiseless and noisy ones corresponding to 1 and 0 capacities.
	
	\subsection{Signal Polarization}
	
	For non-binary real-number constraints, e.g., constellation modulation, MIMO, non-orthogonal multiple access (NOMA), etc., the split data streams also show reliability differentiation. This SP phenomenon comes from natural signal features, e.g., Euclidean distance distinction among modulation bits, diverse propagation channel qualities in MIMO/NOMA systems, different levels of interference among MIMO/NOMA data streams, etc. Fig.~\ref{fig_1} shows the SP phenomenon in the 4-ary pulse amplitude modulation (4PAM), where the minimum Euclidean distance between the first bits differs from that between the second bits, leading to capacity distinction. Nevertheless, limited by signal dimensions and non-binary constraints, SP cannot create the sharp ``Matthew effect (the good gets better and the bad gets worse)'' \cite{KaiNiu} among subchannel capacities as in the CP.
	
	\subsection{Generalized Polarization Transform}
	
\begin{figure}[htbp]
		\centering{\includegraphics[scale=0.28]{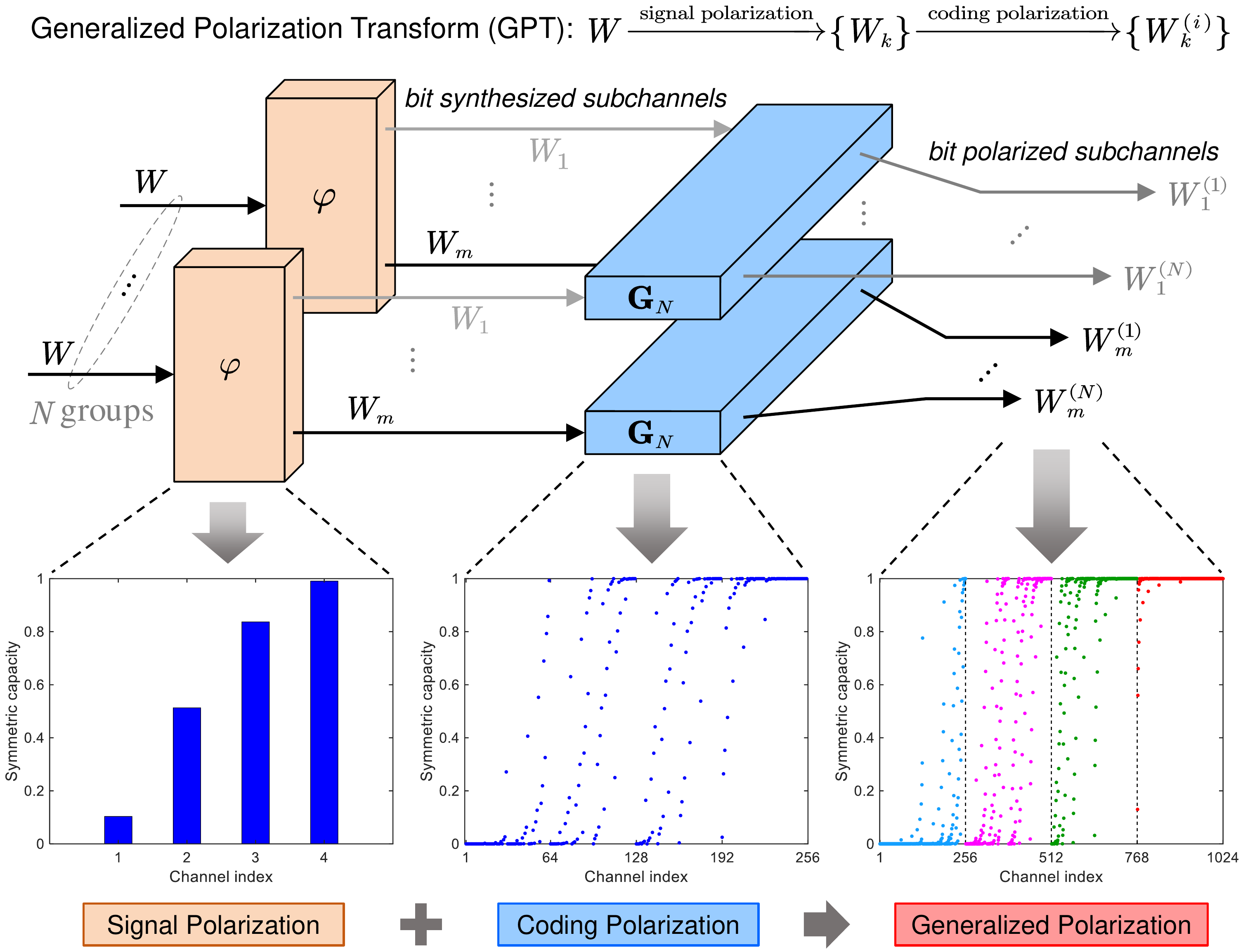}}
		\caption{A diagram of GPT.}
		\label{fig_2}
	\end{figure}

	In this part, we introduce the generalized polarization transform (GPT) to describe CP and SP jointly. The whole structure polarizes the original channel $W$ into a group of good and bad subchannels. See the polar-coded 4PAM example in Fig.~\ref{fig_1}. From right to left, the raw physical channel $W$ is decoupled into $W_1$ and $W_2$ in the SP step, and then CP performs on $W_1$ and $W_2$ to generate the final bit polarized channels $W_1^{\left(1\right)}$, $W_1^{\left(2\right)}$, $W_2^{\left(1\right)}$, and $W_2^{\left(2\right)}$, which have varying capacities. Fig.~\ref{fig_2} goes a step further. The underlying channel $W$ carrying one $2^m$-ary symbol is first transformed to an ordered set of \emph{bit synthesized subchannels} under an order-$m$ SP. According to the chain rule of mutual information, each resulting subchannel is assumed to know the original channel output and the transmitted bits over bit synthesized subchannels with smaller indices. The second stage performs the $N$-dimensional binary CP transform on each of these $m$ bit synthesized subchannels respectively, resulting in $mN$ bit polarized subchannels finally. Observing the distribution of good/bad channels (see the bottom of Fig. \ref{fig_2}), we find that the SP-CP cascaded structure of GPT results in a two-fold Matthew effect: SP preliminarily divides the bit synthesized subchannels into the good and bad ones. Then by CP, those good subchannels create more good bit polarized subchannels and bad subchannels creat more bad ones.
	
	When the blocklength tends to infinity, similar to polar codes, the capacity-achieving property of GPT can also be proven \cite{Seidl,Dai_mimo}. For finite blocklengths, $W$ cannot be fully polarized under GPT. Even so, GPT is adequate for eMBB and URLLC cases. An integrated design of coded transmission system under the GPT paradigm makes full use of the underlying polarization phenomenon inside various air interface modules, thus can be highly efficient in attaining a Shannon-limit-approaching performance under finite blocklengths. For eMBB services that require high data rates and throughput, high-dimensional signals along with long blocklengths under GPT can achieve an intense polarization effect to guarantee high performance. For URLLC services where short-to-moderate-packet traffic dominates, the merit of capacity-preserving and the performance superiority of short polar codes still allow GPT to ensure highly reliable transmissions.
	
	
	\section{GPT Transceiver Design}\label{section_transceiver}
	
	In practice, we can almost continuously adjust the coding rate in the GPT-based transmission systems by adding or deleting one bit subchannel. Following the two-stage structure of GPT, the code construction can be realized robustly and efficiently like the sequence construction in plain polar codes. At the receiver, the SC mechanism embraces detection and decoding as a whole framework.

	\subsection{Transmitter: Encoding and Code Construction}
	
	As aforementioned, the GPT transmission systems are generally formulated with the multilevel polar coding structure. Given the number of coupled data streams $m$ and the blocklength $N $, each component $N$-length polar code corresponds to one bit synthesized subchannel. The binary source block contains $K$ information bits and $mN - K$ frozen bits such that the overall coding rate $R$ is $K/{(mN)}$.
	
	\begin{figure}[htbp]
		\centering{\includegraphics[scale=0.3]{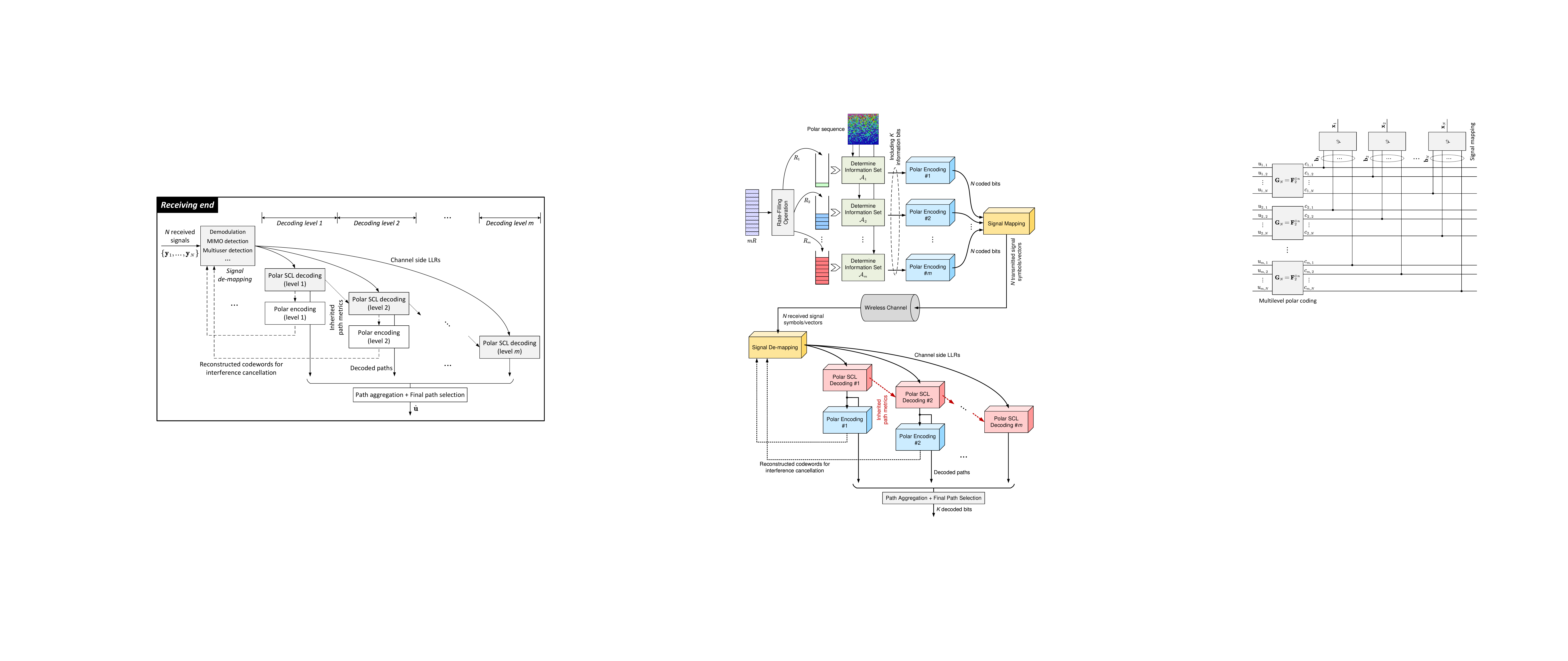}}
		\caption{Transceiver procedures of GPT coded transmission systems.}
		\label{fig_3}
	\end{figure}
	
	Similar to plain polar codes, the reliability of each subchannel after GPT is different, so picking the most $K$ reliable subchannels out of the entire $mN$ ones for carrying information bits is a critical step to minimize the block error rate (BLER) upper bound under SC decoding. In GPT systems, with emphasis, all component polar codes should be treated as a whole for channel selection, i.e., code construction.
	
	Classical online polar code construction methods, like density evolution, Gaussian approximation, have high accuracy but suffer from high computational complexity. Worse still, these methods introduce sorting complexity. Inverse to online code construction, the polarization weight (PW) and the Polar sequence of 5G construct polar codes with predetermined order of channel indices \cite{Bioglio_polar}. However, these offline methods can only be applied to a sole component polar code. To take advantage of the PW/Polar sequence in GPT multilevel systems, one can accurately allocate the number of information bits for each component polar code. Then each information indices set is obtained in a straightforward way. Therefore, the critical issue becomes how to assign each component code rate agilely. Fortunately, this can be tackled by a rate-filling method as follows. First, the equivalent channel whose capacity equals the target transmission rate $mR$ is utilized to determine each component code rate as its corresponding bit subchannel's capacity under SP. Then each information indices set is directly extracted with the aid of PW/Polar sequence with $O(1)$ complexity. This method requires only a one-shot operation demonstrated in Fig. \ref{fig_3}, which answers the fast construction question in practice.
	
\subsection{Receiver: Detection and Decoding}
	
	Recall that the SP-CP cascaded structure introduces the correlation among source bits, so each bit at a given index relies on all of its preceding bits with smaller indices within each component polar codes. This kind of correlation introduced by CP can be conceptually understood as interference in the coding domain. Also, the correlation among data streams split under SP can be treated as interference in the signal domain. SC of the ``interference'' caused by previous bits and data streams can benefit the reliability in source information retrieval. Hence, in GPT systems, the SC operation is further extended to both polar decoding and data stream detection, forming a compound SC structure.
	
	As an enhanced version of SC, the SC list (SCL) algorithm \cite{KaiNiu} can also be used in GPT transmission systems. However, the compound SCL detection and decoding here show some differences from conventional polar decoding. It presents the multilevel structure bridging component polar codes connected by the real-number signal processing module. The conventional SCL decoding with only hard information delivered will cause the correct path to be lost. In consequence, the performance superiority of SCL or cyclic redundancy check (CRC) aided SCL (CA-SCL) decoding \cite{niukai_CASCL} versus SC decoding will be greatly decreased. In summary, the standard list decoding cannot provide satisfactory performance on these GPT coded transmission systems. To tackle this, we have proposed a modified SCL algorithm to integrate the decoding of multiple component polar codes into a collaborative ensemble. The key idea is not only exploiting hard decision information given by past decoded component polar codes but also inheriting soft path metrics to aid the subsequent decoding. By this means, the path metrics are used as the \emph{a priori} information to ensure the correct path being well preserved in the list when the decoding bridges the adjacent component polar codes. The receiver in Fig. \ref{fig_3} demonstrates the procedure of this path metric inherited SCL (PMI-SCL) decoding algorithm.
	
	Despite many salient advantages of GPT, one drawback is the relatively low throughput and parallelism compared to Turbo/LDPC schemes. It is incurred by the intrinsically successive characteristics of decoding and detection. However, the computational complexity of GPT is usually lower than Turbo/LDPC schemes. Even with the above PMI-SCL decoding, it only imposes slight burdens on storing path metrics. Overall, the computational complexity of PMI-SCL is pretty close to SCL decoding.

	\section{Open Issues}
	
	We have demonstrated the benefits of integrated design under the GPT paradigm. Now, we discuss several research directions for further potential improvement in future GPT-based transmissions.
	
	\emph{Constellation shaping for polar-coded modulation:} Regular constellation shaping methods aim at generating Gaussian-like signal points to attain a rate close to the Shannon limit. However, taking the polarization principle into account, the shaping method under GPT should maximize the reliability distinction among bit synthesized subchannels while increasing in the achievable sum rate. It can be realized by optimizing the constellation mapping rule at the SP stage.
	
	\emph{Transmit precoding design of polar-coded MIMO:} Guided by GPT, the transmit precoding design of polar-coded MIMO should follow the polarization criterion. Concretely, it is a two-step process. A basic precoding matrix is first selected for maximizing the capacity. Then it should be post-multiplied by a unitary matrix to maximize the polarization effect among spatial subchannels without eroding the optimized capacity.
	
	\emph{Multiple access mechanism optimization under GPT:} As previous results show, polar-coded multiple access is still in its infancy. There are many open theoretical and practical problems yet to be addressed, especially for the recent developed random access cases. Under the GPT paradigm, polar coding should cooperate with random access protocols to approach its theoretical bound.
	
	\section{Conclusions}
	
	We have proposed GPT as a new paradigm to implement a real integration of air interface modules, envisioning a breakthrough of SE for future wireless communications. We have presented brief tutorials on the GPT principle, GPT transceiver design and reviewed three typical GPT-coded transmission solutions to shed light on the potential of GPT. We believe that GPT can serve the future wireless communication requirements as a promising and viable candidate.
	

	\ifCLASSOPTIONcaptionsoff
	\newpage
	\fi
	


\end{document}